\documentclass[journal]{IEEEtran}
\usepackage{amsmath,amssymb}
\usepackage[utf8]{inputenc}
\usepackage{graphicx}
\usepackage{xcolor}
\usepackage{cite}
\usepackage{keyval}
\usepackage[linesnumbered,ruled,vlined]{algorithm2e}

\newcommand\norm[1]{\left\lVert#1\right\rVert}

\ifCLASSINFOpdf
\else
\fi
\usepackage{algorithmic}

\usepackage[font=small,skip=5pt]{caption}
\begin{document}
 \pagenumbering{gobble}   
\bstctlcite{IEEEexample:BSTcontrol}
%
\title{Joint Channel Estimation and Device Activity Detection in Heterogeneous Networks}

  \vspace{1mm}
\author{Leatile Marata\IEEEauthorrefmark{1}\IEEEauthorrefmark{2}, Onel Luis Alcaraz López\IEEEauthorrefmark{1}, Eduardo Noboro Tominaga\IEEEauthorrefmark{1}, Hirley Alves\IEEEauthorrefmark{1}\\
    \IEEEauthorblockA{
		\IEEEauthorrefmark{1}6G Flagship, Centre for Wireless Communications (CWC), University of Oulu, Finland\\
		\IEEEauthorrefmark{2}Botswana International University of Science and Technology (BIUST), Botswana\\
		\{firstname.lastname\}@oulu.fi\\		
	}    

\thanks{~~}
}

%
%

\markboth{~~}%
{Shell \MakeLowercase{\textit{et al.}}: Bare Demo of IEEEtran.cls for IEEE Journals}
%



\maketitle
\begin{abstract} Internet of Things (IoT) has triggered a rapid increase in the number of connected devices and new use cases of wireless communications. To meet the new demands, the fifth generation (5G) of wireless communication systems features native machine type communication (MTC) services in addition to traditional human type communication (HTC) services. Some of the main challenges are the heterogeneous requirements and the sporadic traffic of massive MTC (mMTC), which makes the orthogonal allocation of resources infeasible. To overcome this problem, grant free non-orthogonal multiple access schemes have been proposed alongside with sparse signal recovery algorithms. While most of the related works have considered only homogeneous networks, we focus on a scenario where an enhanced mobile broadband (eMBB) device and multiple MTC devices share the same radio resources. We exploit the approximate message passing (AMP) algorithm for joint device activity detection and channel estimation of MTC devices in the presence of interference from eMBB, and evaluate the system performance in terms of receiver operating characteristics (ROC) and channel estimation errors. Moreover, we also propose two new pilot sequence generation strategies which improve the detection capabilities of the MTC receiver without affecting the eMBB service.  
\end{abstract}


%
\IEEEpeerreviewmaketitle

\section{Introduction}
%
%
%
%

\IEEEPARstart{T}{he} fifth generation (5G) of wireless communication systems is the first generation that natively features machine type communication (MTC) services in addition to the traditional human type communication (HTC) services. More specifically, 5G features three generic services: enhanced mobile broadband (eMBB) provides very high data rates with high availability for HTC applications, ultra-reliable low-latency communications (URLLC) aims at MTC applications with very stringent latency and reliability requirements, while massive MTC (mMTC) provides massive connectivity to low complexity devices. The latter is seen as one of the key enablers of the Internet of Things (IoT) paradigm \cite{osseiran2014scenarios,dawy2016toward,mahmood2019six}.

\par MTC and HTC services often coexist in the same cellular network, which complicates the network design. In contrast with HTC, MTC is characterized by uplink driven traffic where a large number of devices sends information at low data rates to a central node such as a base station (BS) \cite{liu2018massive2}. To meet the different data rate, reliability and latency requirements imposed by the heterogeneous services, network resources have to be optimally shared among all the coexisting devices. 

\par Previous generations of wireless communication systems relied mostly on orthogonal multiple access and resource allocation techniques \cite{chung2005signaling}. However, such techniques become infeasible as the number of users grows large, which is the case of future mMTC scenarios where the number of devices may be in the order of $10^4-10^6$ devices per cell \cite{mahmood2019six}. Resource allocation is even more challenging when heterogeneous services such as eMBB and mMTC coexist. In contrast with eMBB services, MTC packets are short, and the activation pattern of MTC devices is mostly uplink driven and sporadic. 

\par A possible solution for the aforementioned problems lies in adopting grant-free non-orthogonal multiple access (NOMA) for sharing radio access network (RAN) resources \cite{senel2018grant},\cite{popovski2019wireless}. Even though NOMA techniques improve spectral efficiency, their major drawback is an increased risk of unresolvable collisions. Meanwhile, compressed sensing and sparse signal recovery algorithms have been proposed to deal with the scenario where a massive number of MTC devices coexist and compete for the RAN resources \cite{chen2018sparse},\cite{liu2018sparse}. However, most of the works have not considered the collision resolution in scenarios where heterogeneous devices coexist.  
\subsection{Related Literature}
 
 \par Chen \textit{et al.} proposed the use of approximate message passing (AMP) algorithm in \cite{chen2018sparse} for both sparse device activity detection and channel estimation. Their work exploited the channel statistics to improve the AMP in a multiple measurement vector (MMV) using the vector denoising minimum mean squared error (MMSE) estimator. They also formulated an analytical expression that relates the  state evolution of the AMP and probabilities of false alarm (PFA) and probability of missed detection (PMD). Meanwhile, Jinyoup Ahn \textit{et al.} proposed an expectation propagation mechanism in \cite{ahn2019ep} for joint device activity detection and channel estimation. Their work imposes some priors on the activity levels indicator by iteratively minimizing the Kullback–Leibler (KL) divergence for an assumed prior. In spite of the remarkable performance, the proposed detection and estimation technique is computationally complex with exponential computational growth as compared to the AMP algorithm, which is generally consistent and computationally cheaper.
 In another work \cite{liu2018massive}, Wei \textit{et al.} proposed a MMSE based denoiser AMP  algorithm for both channel estimation and sparse activity detection in a mMTC scenario. Also, a NOMA scheme, where the number of devices is greater than the number of  antennas at the BS, is considered by imposing some scheduling after the active devices have been identified. However, their proposed algorithm is complex since it requires large matrix inversions. Senel \textit{et al.} proposed a non-coherent device activity detection algorithm in \cite{senel2018grant} to improve latency and reliability performance for mission critical MTC. The algorithm uses the denoiser function proposed in \cite{liu2018massive}, thus it is a Bayesian framework.
 \par The coexistence between MTC and eMBB has been studied in many works, e.g. \cite{popovski2019wireless}, \cite{popovski2018}. In \cite{popovski2018}, the authors introduce a communication-theoretic framework for the coexistence of the three 5G services in the uplink of the same RAN, but being limited to single antenna devices and BS. Finally,  \cite{tominaga2021_1,tominaga2021_2} compare orthogonal and non-orthogonal resource allocation strategies for coexistence scenarios between eMBB and critical MTC, and between eMBB and MTC, respectively. It can be noted that, in general, the aforementioned works have not dealt with the issue of MTC activity detection in the presence of interference from other services, which is a problem worth addressing.
 
 \subsection{Contributions of the Paper}
 
 \par Our contributions in this work are two-fold: i) we exploit the AMP algorithm to detect the active MTC devices and estimate their channels in the presence of eMBB traffic, while evaluating the system performance in terms of the receiver operating characteristics (ROC) and channel estimation errors, and ii) we propose two new pilot design strategies suitable for heterogeneous networks that outperform current literature in terms of PMD and PFA, with one of them specifically aiming at reducing the unavoidable non-orthogonality. Illustrated numerical results show a relative root mean square error (RRMSE) improvement of more than $50\%$ when using these pilots. 
 


\par \textbf{Notation:} Boldface lowercase and boldface uppercase letters denote column vectors and matrices, respectively. For instance, $a_i$ is the $i$-th element of vector $\textbf{a}$, and $\mathbf{A}_i$ is the $i$-th row of matrix $\textbf{A}$. $\mathbf{A}_{i,j}$ is the $i$-th row and $j$-th column of matrix $\mathbf{A}$. The superscripts $(\cdot)^*$, $(\cdot)^T$ and $(\cdot)^H$ denote the conjugate,  the transpose and conjugate transpose operations. The magnitude of a scalar quantity or the determinant of a matrix is denoted by $|\cdot|$. We denote the circularly symmetric complex Gaussian distribution with mean $\mathbf{a}$ and covariance $\mathbf{B}$ by $\mathcal{CN}(\mathbf{a},\mathbf{B})$. $\binom{a}{b}$ is the binomial coefficient, while $\mathbb{E}[\cdot]$ is the expectation operator.  

\section{System Model}\label{formulate}

\begin{figure}[t!]
    \centering
    \includegraphics[scale=0.8]{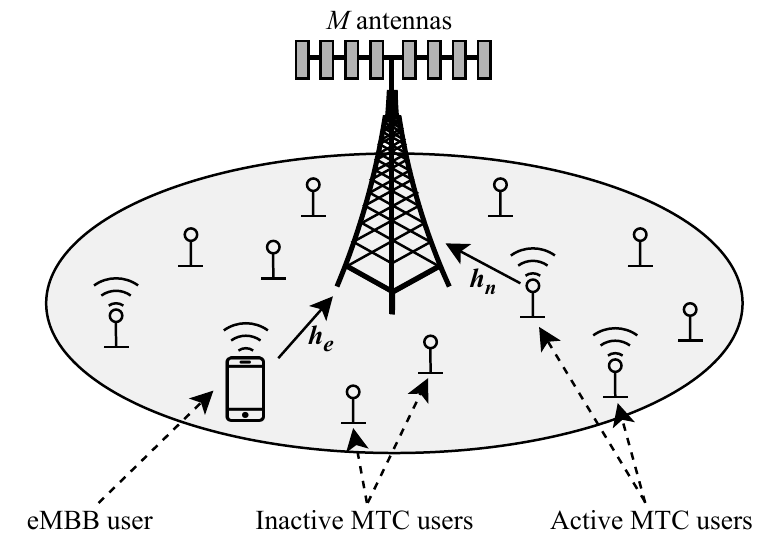}
    \caption{The considered scenario where an eMBB and multiple MTC devices communicate in the uplink to a common BS. In this illustration, only $K=3$ out of $N=11$ MTC devices are active.}
    \label{figMTC2}
    \vspace{0.5cm}
    \includegraphics[scale=0.375]{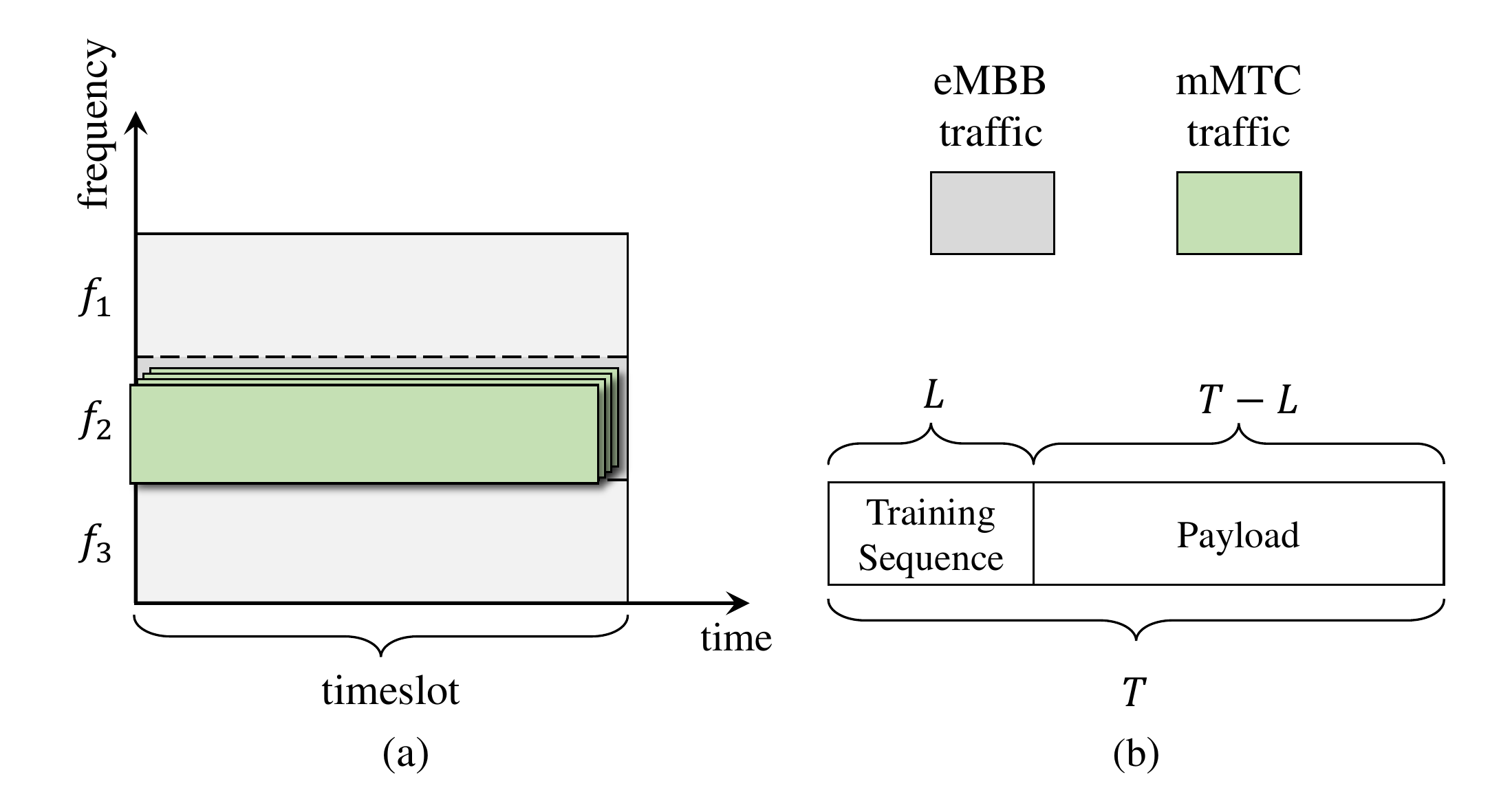}
    \caption{(a) Time-frequency grid for the considered scenario, where one of the frequency channels ($f_2$) is shared by the eMBB and mMTC traffics, and (b) two-phase transmit protocol.}
    \label{Transmission blocks}
\end{figure}

We consider the uplink scenario depicted in Fig.~\ref{figMTC2}, where a single-antenna eMBB device and a set $\mathcal{N} = \{1,2,...,N\}$ of single-antenna MTC devices are in the coverage area of a BS equipped with $M$ antennas. Only $K$ out of $N$ MTC devices are active in a given coherence time interval of $T$ symbols. The channel coefficients between each device and the BS is denoted by $\textbf{h}_i,\;i\in\left\{e,n\right\}$, where the subscripts $e$ and $n$ correspond to the eMBB and the $n$-th MTC device, respectively. The channels are assumed to undergo i.i.d quasi-static Rayleigh fading, i.e., $\mathbf{h}_i\sim \mathcal{CN}(\textbf{0},\beta _i\mathbf{I})$, where $\beta_i$ models the path loss.

\par The eMBB and MTC devices share a common radio resource that is composed of one time slot in a single frequency channel as illustrated in Fig.~\ref{Transmission blocks}a.\footnote{In \cite{tominaga2021_1}, the authors showed that the non-orthogonal resource allocation for eMBB and mMTC outperforms the orthogonal counterpart as the number of antennas at the BS increases.} The eMBB user is assumed to occupy this RAN resource during a long period of time \cite{popovski2019wireless}. On the other hand, each MTC device is intermittently active in each coherence time with probability $\epsilon$, thus there are $K=\epsilon N $ active MTC devices on average in every coherence time interval. We define the activity indicator function for each MTC device as follows
\begin{equation}
  \alpha_n= 
  \begin{cases} 
      1 ,~~\text{if device $n\in\mathcal{N}$ is active} \\
      0 ,~~\text{otherwise}
   \end{cases}.
\end{equation}
All the active devices send training pilots, with equal transmission power $\rho_{u}$ to the BS in each coherence interval such that the BS can perform  CSI estimation, and also activity detection in case of the MTC devices. Let $L$ be the number of pilot symbols transmitted by every active device during a coherence time interval comprising $T$ symbols. Then, the payload can only be transmitted in the remaining $T-L$ symbols, thus $L<T$ is required, as illustrated in Fig. \ref{Transmission blocks}b. One important characteristic of the considered scenario is that the number of devices is usually very large, i.e., $N\gg L$, which can result in collisions owing to the  limited number of orthogonal pilot sequences. Meanwhile, pilot contamination is unavoidable when adopting non-orthogonal pilot sequences, and must be carefully considered. The structure of the pilots plays a key role in the successful joint detection and channel estimation of devices in this heterogeneous NOMA (H-NOMA) system.      
     
\subsection{Pilot Sequences Design}

\par We first define the matrix $\mathbf{A}=[\mathbf{a}_1,\mathbf{a}_2, \cdots,\mathbf{a}_N]$, whose columns are the different pilot sequences of length $L$ adopted by the MTC devices. Thus, the pilot sequence adopted by the $n$-th MTC device is $\mathbf{a}_n =[a_{n,1},a_{n,2},\cdots,a_{n,L}]^T$. The eMBB device is also allocated a pilot sequence of same length denoted by $\mathbf{a}_e$. Each pilot sequence has a unit norm, i.e., $\norm{\mathbf{a}}_{2}^{2}=1$, to ensure that $\mathbf{A}$ has the restricted isometric property to promote sparse recovery when using the AMP algorithm \cite{liu2018massive}. 
  
\par The pilot sequences are generated as follows. First, $L$ orthogonal pilot sequences $ \mathbf{V} =[\mathbf{v}_1,
\mathbf{v}_2,\cdots,\mathbf{v}_{\substack{L}}]$, are generated, e.g., by using a Hadamard pilot matrix \cite{cheng2020orthogonal}. Then, one sequence $\mathbf{v}_i$ is selected and allocated to the eMBB device, i.e., $\mathbf{a}_e=\mathbf{v}_i$, while the remaining $L-1$ orthogonal sequences are then linearly combined in $N$ different ways to yield $N$ non-orthogonal pilots for the MTC devices, i.e., 
\begin{equation}
    \mathbf{a}_n =\sum_{\substack{j=1, j\neq i}}^{L}\vartheta_j \mathbf{v}_{j},  \forall n\in\mathcal{N},
\end{equation}
 where $\vartheta_j$ corresponds to scalar weights and normalized such that $\norm{\mathbf{a}_n}_{2}^2=1,~ \forall n\in\mathcal{N}$. Note that $\mathbf{a}_n^H\mathbf{a}_e=0, ~\forall n\in\mathcal{N}$, while $\mathbf{a}_n^H\mathbf{a}_m\neq0,~  \forall n,m\in\mathcal{N} $
and arbitrary weights $\vartheta$. In other words, the pilot sequence assigned to the eMBB device is orthogonal to all of the pilot sequences assigned to the MTC devices, whereas any two sequences assigned to MTC devices are non-orthogonal. Pilots are assumed to be pre-assigned in order to facilitate the user identification.

\par In this work, we assess the impact of different pilot sequence design strategies. We propose two strategies: Proposed Pilot I and Proposed Pilot II. For Proposed Pilot I, we randomly pick a minimum number of columns $z$ of $\mathbf{V}$ that can be combined with different weights to form $N$ MTC pilot sequences, while maintaining low probability of collision i.e., $\frac{1}{\binom{L-1}{z}}\leq \xi$, where $\xi$ is a very small number. On the other hand the Proposed Pilot II is designed by combining all columns of $\mathbf{V}$ by using equal weights.
 
 \subsection{Signal model}
 The composite signal $\mathbf{Y}\in \mathbb{C}^{L\times M}$ received at the BS as a result of the pilot training phase is given by
 \begin{equation}
\mathbf{Y} = \mathbf{a}_{e}\mathbf{h}_e^T + \sum_{n=1}^{N}\alpha_{n}\mathbf{a}_{n}\mathbf{h}_{n}^T +\mathbf{W} =
 \mathbf{Q}_e+ \mathbf{A}\mathbf{X}+\mathbf{W},
\label{E21}
\end{equation}
where $\mathbf{Q}_e=\mathbf{a}_{e}\mathbf{h}_e^T  \in \mathbb{C}^{L\times M}$ is the received pilot signal from the eMBB device, $\mathbf{A} \in \mathbb{C}^{L\times N}$ is the pilot matrix of the MTC devices, $\mathbf{X}= [\mathbf{x}_1,\mathbf{x}_2,\cdots,\mathbf{x}_N]^T\in \mathbb{C}^{N\times M}$, composed of rows $\mathbf{x}_n=\alpha_n\mathbf{h}_n^T$, corresponds to the effective channels from the MTC devices to the BS, and $\mathbf{W}\in \mathbb{C}^{L\times M}$, composed of columns $\mathbf{w}\sim\mathcal{CN}(\mathbf{0}_{L\times 1},\sigma^2\mathbf{I}_{L\times L})$, is the matrix containing the additive white gaussian noise (AWGN) samples. It is worth noting that matrix $\mathbf{X}$ is sparse along its rows, which facilitates the detection of active MTC devices by recovery of the non-zero rows.

\par Since eMBB  and MTC devices share the same radio resource, we first aim to estimate the eMBB channel $\mathbf{h}_e$, and consequently $\mathbf{Q}_e$. This is recommended since the eMBB device is known to be active, e.g., until a service termination procedure is executed. After decoding the signal from the eMBB device, the BS removes its corresponding contribution $\mathbf{Q}_e$ from the composite receive signal $\mathbf{Y}$ in (\ref{E21}) via successive interference cancellation (SIC). Then, the BS attempts to detect the active MTC devices and estimate their channels by estimating $\alpha$ and $\mathbf{h}$, respectively. After the joint detection and channel estimation procedures, the BS proceeds to the coherent decoding of the remaining $T-L$ symbols in the data transmission phase. 
\section{Joint User Detection and Channel Estimation }\label{label}
Since the eMBB device has been granted a pilot sequence that is orthogonal to all of the pilot sequences assigned to the MTC devices, its corresponding channel is first estimated using the MMSE approach. Specifically, the BS multiplies the received signal by the conjugate of the known eMBB pilot signal to obtain
\begin{equation}
    \mathbf{y} = \mathbf{Y}\mathbf{a}^{*} =L \mathbf{h}_e + \mathbf{W}\mathbf{a}^{*}. 
\end{equation}
 Note that the pilot signals from the MTC devices do not interfere with the eMBB pilots because they are orthogonal to each other. Thus, CSI estimation of the eMBB channel is not affected  and the uplink channel estimate $\mathbf{\hat{h}}_{e}$ is  \cite{ozdogan2019massive}
\begin{align}
  & \hat{\mathbf{h}}_e = \mathbf{\beta}_e\mathbf{\Psi}\mathbf{y}\\
 & \mathbf{\Psi} =L \text{Cov}\{\mathbf{h}_e + \mathbf{W}\mathbf{a}^{*}\}^{-1} = \frac{1}{L\beta_e + \sigma^2}\mathbf{I}. 
 \end{align}

To characterize the distribution of the estimate, we write the error covariance matrix as 
\begin{equation}
    \mathbf{C}_e = \mathbf{\beta}_e\mathbf{I} - L\beta_e^2\mathbf{\Psi},
\end{equation}
such that the MMSE and the error distributions are given by 
\begin{align}
  \hat{\mathbf{h}}&\sim \mathcal{CN}(\mathbf{0}_{M\times 1},\mathbf{\beta}_e\mathbf{I}-\mathbf{C}_e),\\ \tilde{\mathbf{h}}&\sim \mathcal{CN}(\mathbf{0}_{M\times 1},\mathbf{C}_e),   
\end{align}
respectively. After computing the CSI estimate, BS performs SIC \cite{ngo2020multi} such that the estimate $\hat{\mathbf{Q}}_e = \mathbf{a}_e\hat{\mathbf{h}}^{T}$ is subtracted from the composite signal. Then, the new resulting composite signal $\breve{\mathbf{Y}}$ is given by 
\begin{equation}
    \label{Y_breve}
   \breve{\mathbf{Y}} = \mathbf{Y} - \hat{\mathbf{Q}}_e.
\end{equation}
It should be noted that $\breve{\mathbf{Y}}$ still has some level of interference remaining from the eMBB signal owing to the imperfect eMBB channel estimate. Thus, (\ref{Y_breve}) can be written as
\begin{equation}
   \breve{\mathbf{Y}} = \mathbf{A}\mathbf{X} + \mathbf{W}_{eq}, 
\end{equation}
where $\mathbf{W}_{eq}\in \mathbb{C}^{L\times M}$ is the sum of the receiver noise and the error from the imperfect eMBB channel estimate. Clearly, the MTC detection performance deteriorates as such noise and estimation errors increase. Note that the columns of $\mathbf{W}_{eq}$ given a certain error estimate $\mathbf{\tilde{h}}$ are distributed as 
\begin{equation}
     \mathbf{w}_{eq}|\mathbf{\tilde{h}}\sim~\mathcal{CN}(\mathbf{\tilde{h}}, \sigma^2\mathbf{I}). 
\end{equation}
To perform the detection of the MTC devices, the estimate $\hat{\mathbf{X}}$ of $\mathbf{X}$ can be approximately solved using an AMP iterative procedure shown in \textbf{Algorithm~\ref{algoAMP}} . The algorithm takes as inputs the matrix $\breve{\mathbf{Y}}$, the acceptable error $\Delta$ and the denoising function, which in this case is the MMSE defined in \cite{senel2018grant,liu2018massive} as
\begin{equation}
    \label{eta}
        \eta(\mathbf{x}) = \frac{\beta_n(\beta_n\mathbf{I}+\mathbf{\Sigma})^{-1}\mathbf{x}}{1+\frac{1-\epsilon}{\epsilon}	\eth(\mathbf{x})|\mathbf{I}+\beta_n\mathbf{\Sigma^{-1}}|},
\end{equation}
where $\eth(\mathbf{x})= \mathrm{exp}(-\mathbf{x}^{H}\mathbf{\Sigma}^{-1}\mathbf{x}+ \mathbf{x}^{H}(\beta_n \mathbf{I}+\mathbf{\Sigma})^{-1}\mathbf{x})$, and $\mathbf{\Sigma}$ is the state evolution of the AMP, which is defined by \cite{liu2018massive}
\begin{multline}
     \mathbf{\Sigma}^{t+1} = \frac{\sigma^2\mathbf{I}}{L\rho_u} + \frac{N}{L}\mathbb{E}\big[\big(\eta(\mathbf{x}_{\beta}+(\Sigma^{t})^{\frac{1}{2}}\mathbf{s})-\mathbf{x}_{\beta}\big)\\\big(\eta(\mathbf{x}_{\beta}+(\Sigma^{t})^{\frac{1}{2}}\mathbf{s})-\mathbf{x}_{\beta}\big)^H
    \big],
  \end{multline}
where $t$ is the iteration index,  $\mathbf{s}\sim\mathcal{CN}(\mathbf{0},\mathbf{I})$ is a random error added to the true channel whose average distribution is $\mathbf{x}_{\beta}\sim (1-\epsilon)\delta_0 +\epsilon p_{h_{\beta}}$, $\delta_0$ corresponding to the zeros of the inactive devices, and $p_{h_{\beta}}$ is the distribution of the channel vector of the active devices. Ideally, the denominator of (\ref{eta})  will either tend to $\infty$ or $1$, thus the output of the denoising function will either be a vector of zero entries $\textbf{0}$ or a vector of non-zeros entries $\hat{\mathbf{x}}_n$. In the former case, devices are declared inactive, while they are declared active in the latter. We initialize the algorithm with $\hat{\mathbf{X}}=\mathbf{0}$ and $\mathbf{R}=\breve{\mathbf{Y}}$ at $t=0$. Then, the iterative procedure between lines 2 and 6 is repeated until the difference between the previous and current residuals fall below the acceptable error $\Delta$ set at $10^{-4}$. In line 4, the rightmost term is the Onsager term \cite{liu2018massive} where  $\langle\mathbf{b} \rangle=\frac{1}{N}\sum_{n=1}^{N}b_n$ and $\eta'(\cdot)$ is the first order derivative of the denoising function. Aiming at a fair analysis, devices are declared active if, for a given threshold $\zeta$,  $\norm{\mathbf{\hat{x}}_n}_2\geq \zeta$, and inactive otherwise. One of the advantages of the AMP algorithm is the fact that once the device is active, the vector $\mathbf{\hat{x}}_n^{t}$ is also its channel estimate. Nevertheless, CSI estimates can still be refined after the detection phase. 
\begin{algorithm}[!t]
  \KwIn{$\breve{\mathbf{Y}}$, $\Delta$, $\eta(\mathbf{x})$}
  \KwOut{$\hat{\mathbf{X}}$}
  Initialisation: $\mathbf{X}^0 = \mathbf{0}$, $\mathbf{R}^0=\breve{\mathbf{Y}}$, $t= 0$\\\
    \Repeat{$|\mathbf{R}^{t+1}-\mathbf{R}^t| < \Delta$}{%
      \label{stepA}
      
   $\mathbf{x}_n^{t+1} = \eta(\mathbf{R}^H\mathbf{A} +\mathbf{x}_n^t)$
   
    $\mathbf{R}^{t+1}=\breve{\mathbf{Y}}-\mathbf{A}\mathbf{X}_{n}^{t+1}+\frac{N}{L}\langle \eta'(\mathbf{R}^H\mathbf{a}_n+\hat{\mathbf{x}}_n^t ) \rangle$\

      
       $t = t + 1$\
    }
  \caption{AMP detection after SIC}
  \label{algoAMP}
\end{algorithm}

\section{Results and Numerical Analysis}\label{results}

\begin{figure}[t!] 
\centering
        \includegraphics[width=0.45\textwidth]{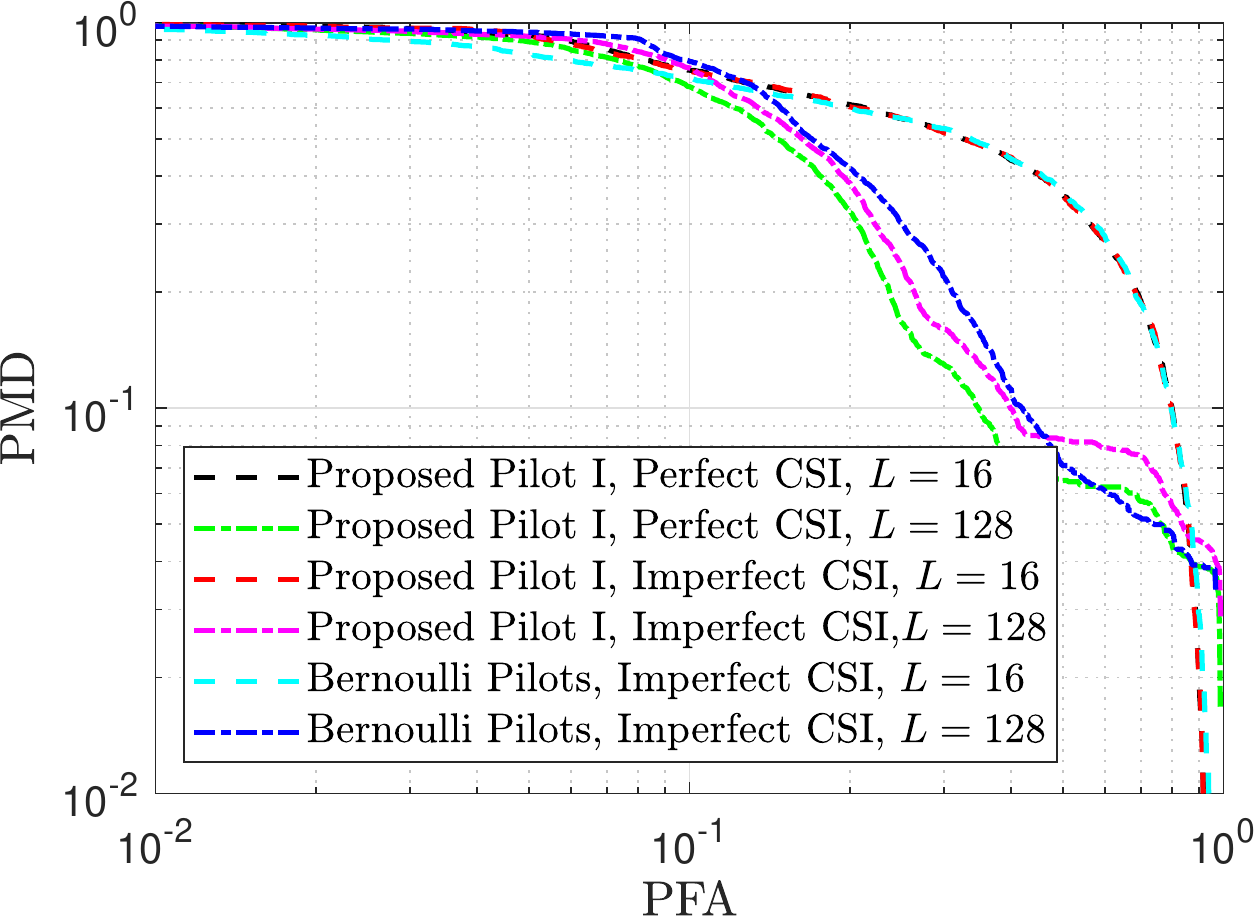}
        \caption{ROC for perfect and imperfect CSI using Bernoulli pilots and Proposed Pilot I.} \label{csiCheck} 
        \vspace{0.75cm}
        \includegraphics[width=0.45\textwidth]{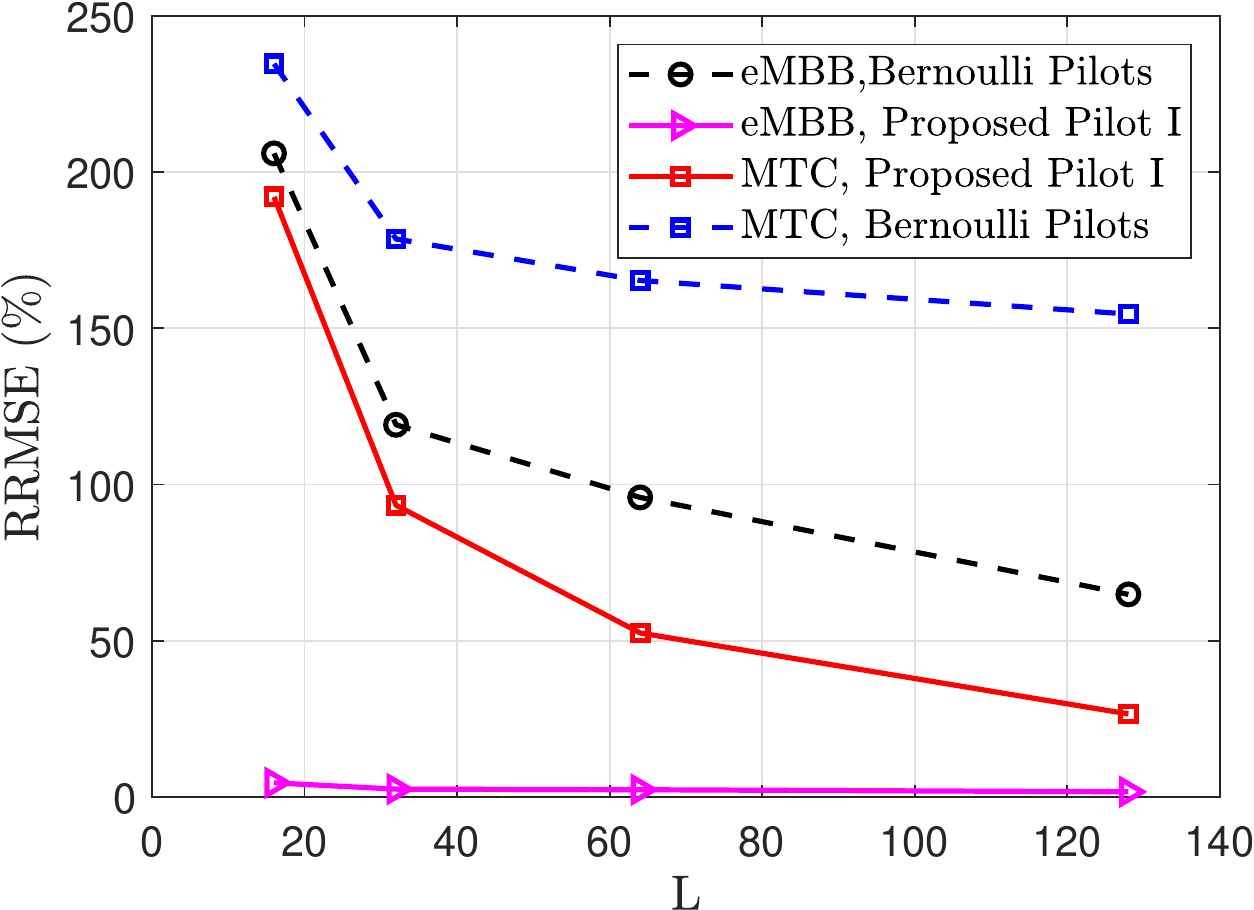}
        \caption{RRMSE for imperfect CSI as a function of the pilot length.} \label{mmseFig} 
\end{figure}

The configuration used for testing the performance of the proposed pilots is $M=20$, $N=200$, $\epsilon=0.05$ and $\xi = 0.001$. Fig.~\ref{csiCheck}  compares the detection performance of the AMP algorithm using the Proposed Pilot I and the Bernoulli pilots from \cite{senel2018grant}. We note that, under imperfect CSI and different values of $L$, the Proposed Pilot I outperforms the Bernoulli Pilots. This happens because it is impossible to completely remove interference from the eMBB device when using Bernoulli pilots due to the non orthogonality among them. To further support our results, Fig. \ref{mmseFig} shows the RRMSE achieved when using the Proposed Pilot I and Bernoulli pilots for a maximum $\mathrm{PMD} = 20\%$. It can be observed that, for $L=128$, the Proposed Pilot I achieves a RRMSE of $1.63\%$ in the estimation of the eMBB channel, while for Bernoulli pilot the RRMSE is of $64.8\%$. As expected, the lower the RRMSE of the eMBB CSI estimate, the better the CSI estimate of the MTC devices, as evidenced by RRMSE of $26.55\%$ and $154.5\%$ for Proposed Pilot I and Bernoulli Pilots respectively at $L = 128$. 

\par
\begin{figure}[t!] 
    \centering
    \includegraphics[width=0.45\textwidth]{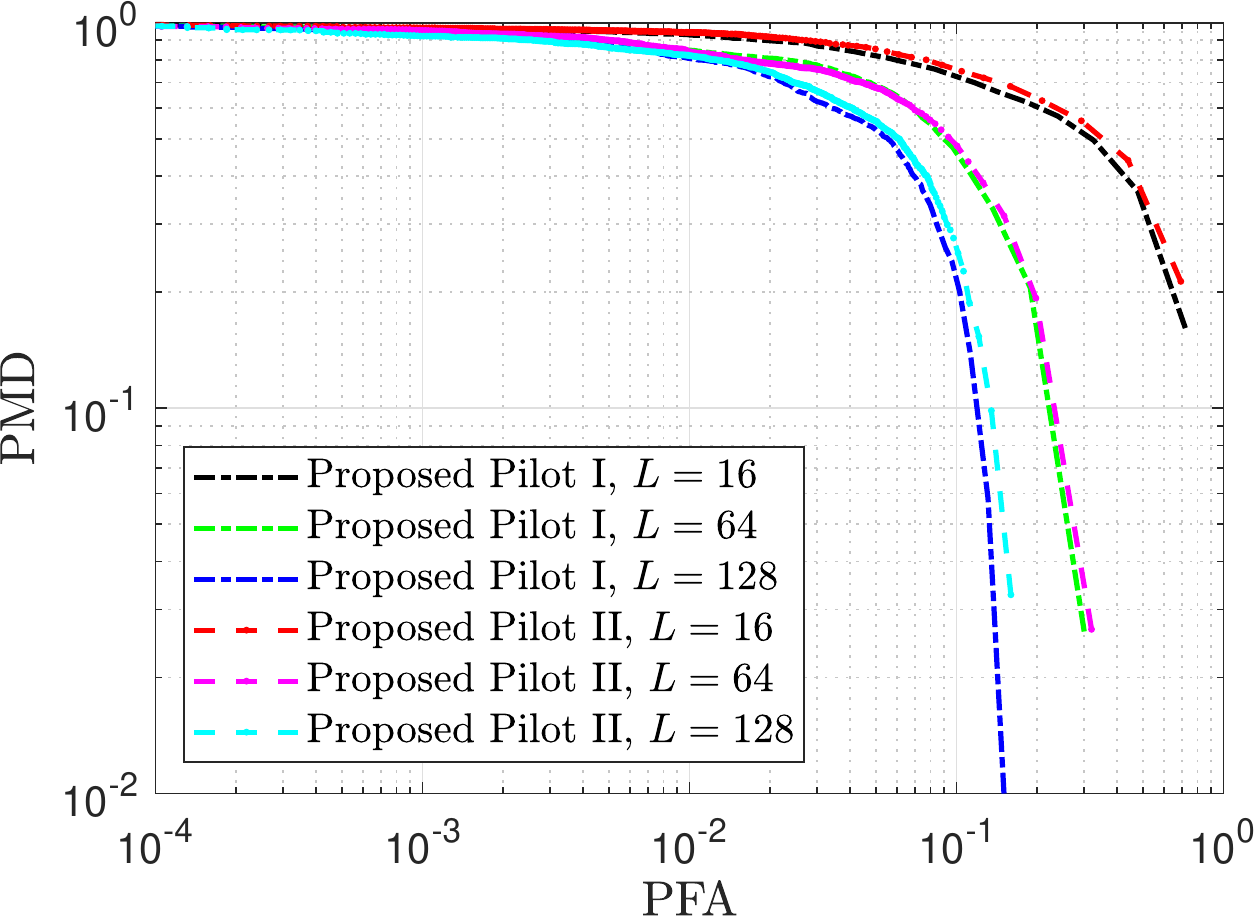}
    \caption{ROC for the proposed pilot generation strategies, for imperfect CSI.} \label{weights} 
    \vspace{0.75cm}
    \includegraphics[width=0.45\textwidth]{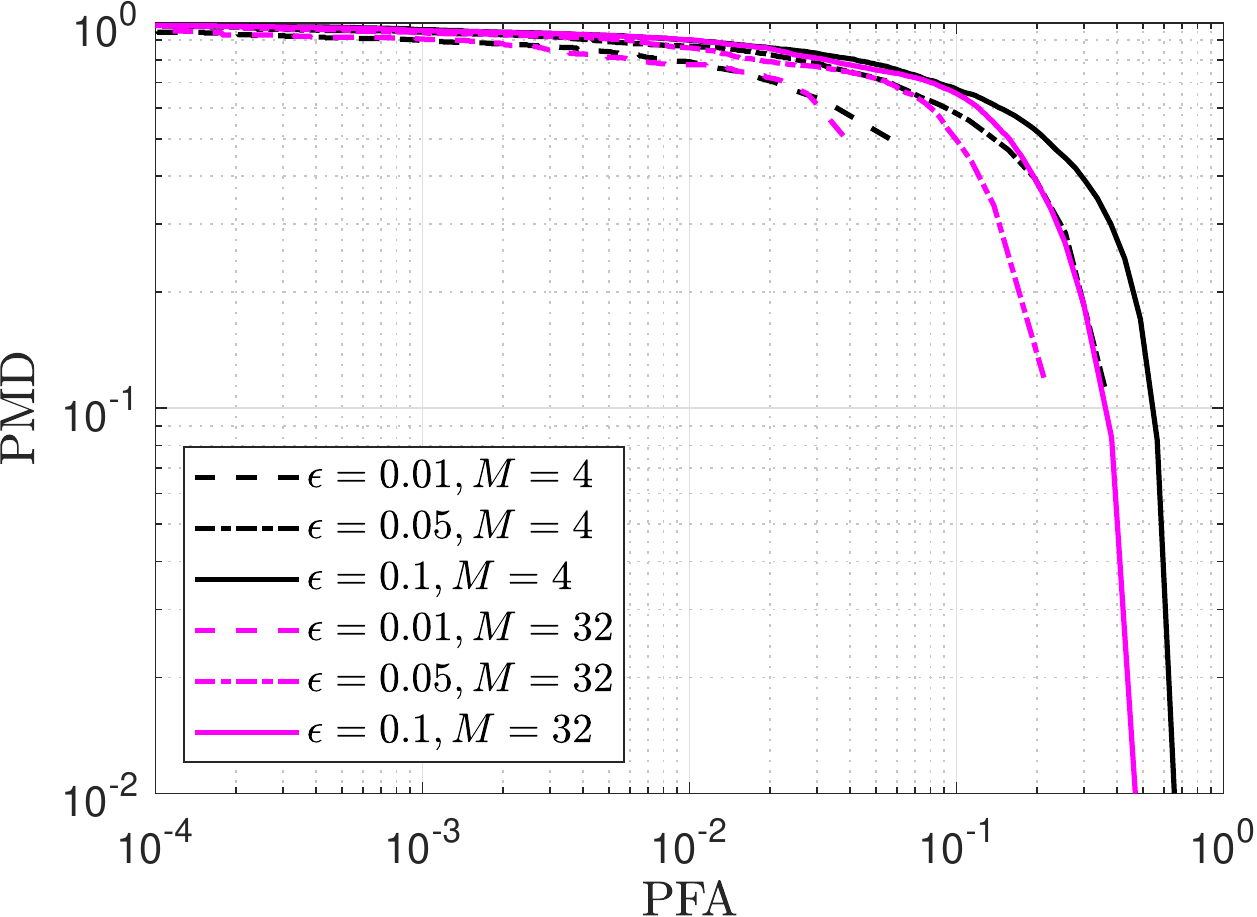}
    \caption{ROC considering imperfect CSI, for $L= 64$, with different sparsity levels and number of antennas using Proposed Pilot I.} \label{Antennas} 
\end{figure}

 As expected and illustrated in Fig.~\ref{weights}, a better detection performance, i.e., lower PFA and PMD, is reachable as the length of the pilots increases. This happens because as $L$ approaches $N$, the pilot sequences tend to be orthogonal to each other. Observe also that as the pilot sequence length increases, the Proposed Pilot I outperforms the Proposed Pilot II as evidenced by the lowest $\mathrm{PMD} = 14.64\%$ for Proposed Pilot I as compared to $\mathrm{PMD} = 32.77\%$ achieved by Proposed Pilot II, both for $\mathrm{PFA} = 17.51\%$. This is due to the fact that the Proposed Pilot I is designed such that it reduces the non-orthogonality more efficiently among the pilot sequences by minimizing the number of non-orthogonal pilots to be used. 
       

\par Finally, Fig. \ref{Antennas} shows the ROC for $L=64$, $N=200$ for different number of receive antennas and sparsity levels. Observe that the detection capability improves as the number of antennas increases, which is consistent with the results from \cite{senel2018grant}. However, as also pointed out by the authors of \cite{senel2018grant}, the performance gains owing to an increasing number of antennas is less significant compared to the performance gains obtained by increasing pilot sequence lengths. It can also be noted from the figure that the AMP algorithm is highly sensitive to the sparsity levels, as can be seen in the extreme case of almost equal ROC for $M=4,\epsilon=0.05$, and $M= 32, \epsilon=0.1$. 

\section{Conclusion and Future Directions}\label{conclude}
In this work, we addressed the problem of joint channel estimation and activity detection in a heterogeneous network where multiple MTC devices share the same radio resources with an eMBB device. The presence of the latter imposes a difficulty for the design of pilot sequences. To address this problem, this work proposed two different pilot sequence generation strategies, with Proposed Pilot I proving to improve detection performance of MTC devices by efficiently reducing the non-orthogonality among the pilots allocated to MTC devices. As in homogeneous network scenarios studied in related works, our results showed that the detection performance of the receiver is improved either by increasing the pilot sequence length or the number of antenna elements at the receiver. Future works can consider schemes for automatic tuning of the sparsity level through machine learning techniques, and pilot generation strategies that can be applied in cases where there is more than one eMBB device.   
\section*{Acknowledgments}
This research has been financially supported by Academy of Finland, 6Genesis Flagship (Grant no318927), EE-IoT (no319008), and FIREMAN (no 326301) and BIUST.


%





\ifCLASSOPTIONcaptionsoff
  \newpage
\fi



\bibliographystyle{IEEEtran}
\bibliography{eurosip.bib}




%



\begin{IEEEbiographynophoto}{~~}
~~
\end{IEEEbiographynophoto}




\end{document}